\newcommand{\ifb}{\ensuremath{\mathrm{fb^{-1}}}}
\newcommand{\ipb}{\ensuremath{\mathrm{pb^{-1}}}}
\newcommand{\TeVcc}{\ensuremath{\,\mathrm{Te\kern -0.1em V\!/c^{2}}}}
\newcommand{\GeVcc}{\ensuremath{\,\mathrm{Ge\kern -0.1em V\!/c^{2}}}}
\newcommand{\MH}{\ensuremath{M_{\mathrm{H}}}}
\newcommand{\MZ}{\ensuremath{M_{\mathrm{Z}}}}
\newcommand{\MW}{\ensuremath{M_{\mathrm{W}}}}
\newcommand{\MA}{\ensuremath{M_{\mathrm{A}}}}

\documentclass[slac_one]{revtex4}
\usepackage{graphicx}
\usepackage{fancyhdr}
\usepackage{subfig}

\begin{document}

\title{Prospects for the Higgs Boson Searches with CMS}

\author{Matteo Sani}
\affiliation{University of California, San Diego}

\begin{abstract}
An overview on the prospects for the Higgs boson searches with the CMS detector is
presented.
Projections have been made to estimate the potential to a possible discovery or 
exclusion of the Higgs boson
during the run at a center of mass energy of 7~$\TeVcc$ at LHC, with a
recorded integrated luminosity of approximately 1~$\ifb$, conditions 
expected by the end of 2011.
\end{abstract}

\maketitle

\thispagestyle{fancy}

\section{CURRENT STATUS OF HIGGS SEARCHES}
At the end of the Large Electron Positron (LEP) operations, an
exclusion limit was reached by combining the results of all the Higgs search
channels performed by its four experiments, establishing a lower
limit on $\MH$ of 114.4~$\GeVcc$ at 95\% confidence level.

More recently, the direct Higgs searches performed by the CDF and D0
experiments at the Tevatron accelerator have reached better
sensitivity, making possible the exclusion of a SM Higgs Boson with a
mass between 158 and 174~$\GeVcc$.


\section{HIGGS PRODUCTION AND DECAY MODES AT THE LHC}

The LHC is continuously collecting luminosity at a center of 
mass energy of 7~$\TeVcc$ since March 2010. 
The projected luminosity to be delivered before the
shut-down to go to higher energies is 10~$\ipb$ by November 2010 and 1~$\ifb$
by the end of 2011.

The main production modes for the Standard Model (SM) Higgs at 7~$\TeVcc$ at
the LHC, presented in Fig.~\ref{production modes}, are: 
gluon gluon fusion (gg), dominant in the whole mass ranges; 
vector boson fusion (VBF, qqH), increasingly
important for high masses and leading to a characteristic signature in
the final state with two forward jets and associated production, with
W and Z boson or top quarks (WZ, ZH, t$\mathrm{\bar{t}}$H), that is relevant in the low
mass region and allows for easy triggering.
\begin{figure*}[t]
\centering
\subfloat[] {
\includegraphics[width=80mm]{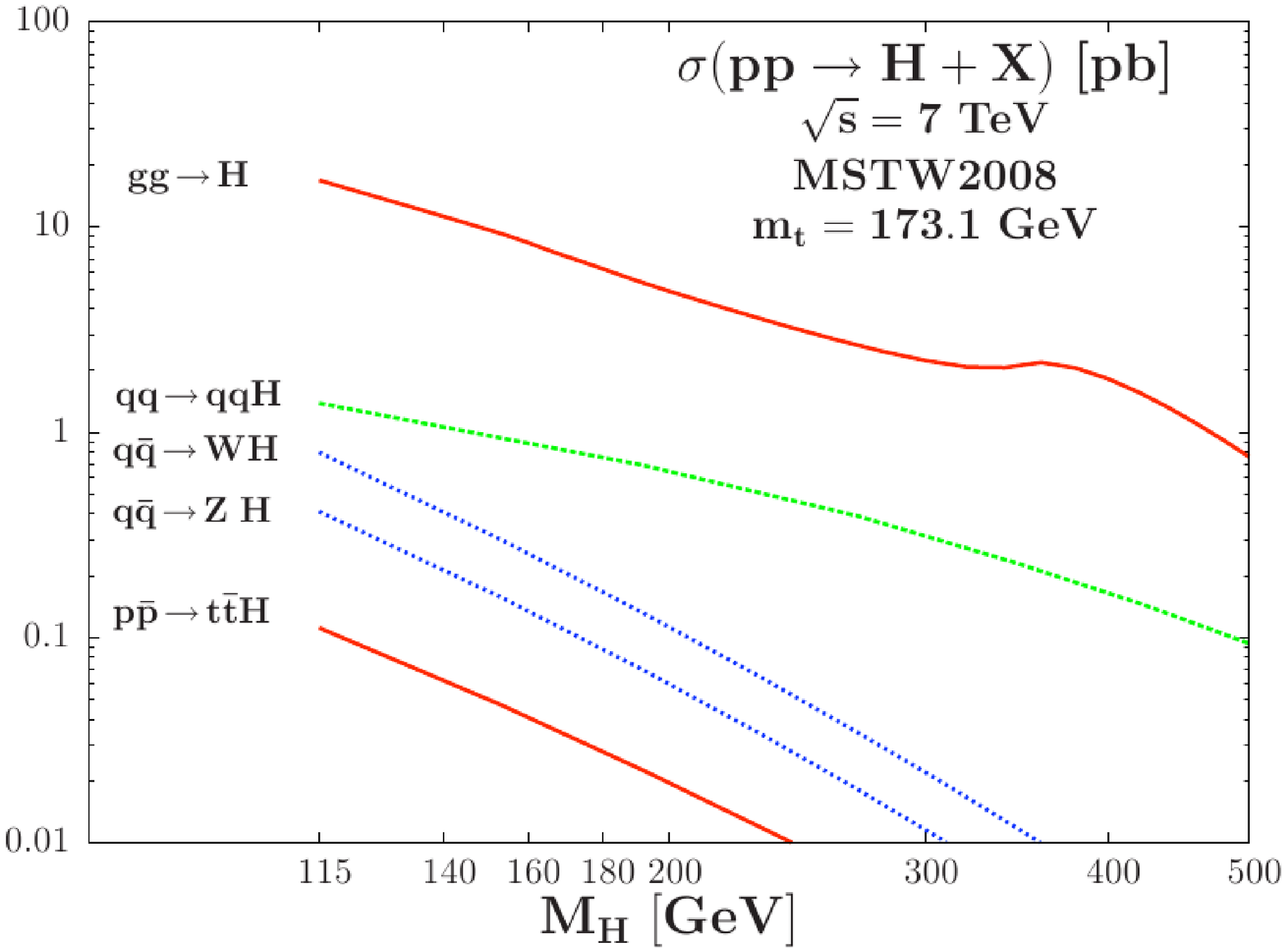}
\label{production modes}
}
\subfloat[]{
\includegraphics[width=90mm]{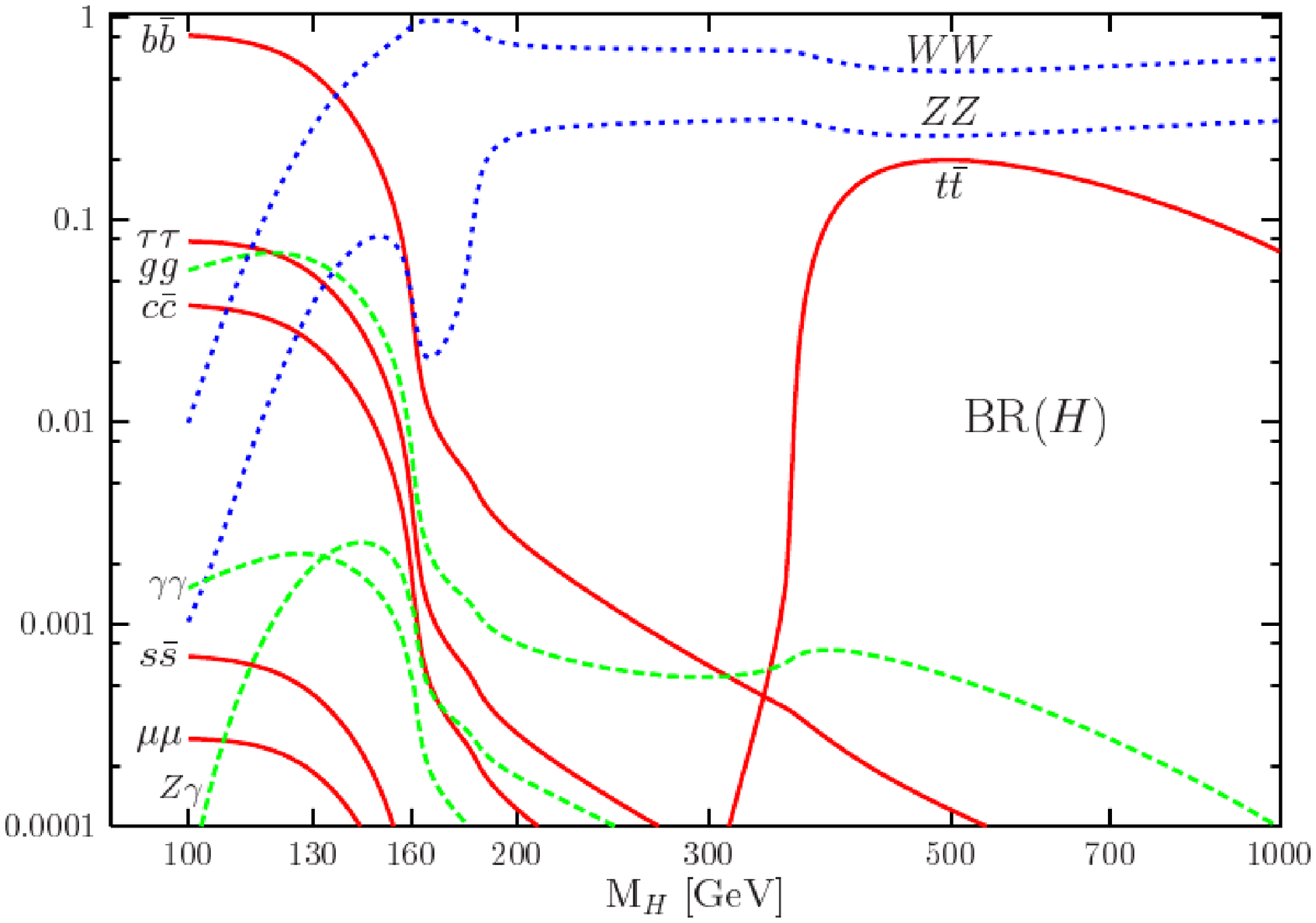}
\label{decay modes}
} 
\caption{Standard Model Higgs production cross sections for a proton
proton collider with 7~$\TeVcc$ center of mass energy (a) and decay modes (b). 
The cross sections of the various processes and the branching ratios are presented
as a function of the Higgs boson mass.} 
\end{figure*}

Concerning the decay modes, presented in Fig.~\ref{decay modes}, 
for low masses ($\MH<2 \MZ$), H$\rightarrow$b$\mathrm{\bar{b}}$ decay has the highest branching ratio, although
this channel is experimentally challenging due to the huge QCD
background present. Another important decay mode for low masses is
H$\rightarrow\tau\tau$, accessible through VBF, and H$\rightarrow\gamma\gamma$, 
in which the Higgs mass peak can be reconstructed with good resolution. 
For high masses ($\MH > 2 \MZ$), the H$\rightarrow$t$\mathrm{\bar{t}}$ decay can be 
studied, however, it presents difficulties in the selection due to the small
signal-to-noise ratio.

The H$\rightarrow$WW$^{*}$ and H$\rightarrow$ZZ$^{*}$ decay channels are powerful in the whole mass
range. The Higgs decay into W bosons gives the earliest sensitivity and
it is the dominant decay mode for a wide mass range. Higgs to ZZ$^{*}$ has
a very clean experimental signature with four leptons where a narrow
mass peak is reconstructed above a smoothly varying background.

\subsection{LHC AND Tevatron Scenarios}
The Higgs searches at CMS during the LHC first period of data taking
will overlap with the last runs of the Tevatron experiments, CDF and D0. 
By the end of 2011, the Tevatron is expected to have 10~$\ifb$ at a center
of mass energy of 1.96~$\TeVcc$ compared with 1~$\ifb$ at 7~$\TeVcc$ for the LHC.

To give a rough estimate of the expected performance of LHC with 
respect to Tevatron the different parton luminosities of the two colliders need to be taken into account.

In the high mass regime ($\MH > 140\GeVcc$) gluon gluon luminosity 
at LHC is more than 15 times higher than at Tevatron.
Being the dominant background to the main channels (H$\rightarrow$ZZ and H$\rightarrow$WW) mainly produced by q$\mathrm{\bar{q}}$ which rises relatively slowly it can be
concluded that LHC will be competitive with 1~$\ifb$.

Contrary in the low mass region the rise in the cross section 
is less pronounced and the q$\mathrm{\bar{q}}$ processes (e.g. Higgs-strahlung) are not 
so favorable at LHC compared to at Tevatron.
The signal to background ratio is also worsen by the dominant 
gluon fusion mechanism which enhances the main backgrounds channels (e.g t$\mathrm{\bar{t}}$, 
W/Zb$\mathrm{\bar{b}}$).
The previous considerations suggest to use the challenging $\gamma\gamma$ mode.

\subsection{Prospective Higgs Searched at CMS}
The CMS experiment utilizes a wide range of Higgs decay channels and
have public results available for different $\sqrt(\textrm{s})$.
The Higgs sensitivity at 7~$\TeVcc$ with 1~$\ifb$, has been
studied using projections made from the results obtained using 
Monte Carlo samples at different center of mass energies, namely 10 and 14~$\TeVcc$.

These projections are not new analysis done with 7~$\TeVcc$ Monte Carlo
samples and new detector simulation and reconstruction software. They
are done starting from public results at 10 and 14~$\TeVcc$.
The signal and background event counts have been re-scaled by the ratio of
7~$\TeVcc$ to 14~$\TeVcc$ cross-sections and then normalized to 1~$\ifb$. No corrections for higher acceptance at smaller $\sqrt(s)$ or for
improvements in the reconstruction are applied.
Systematic and statistical uncertainties have been also re-scaled conservatively.

\section{SM HIGGS SEARCH CHANNELS AT THE LHC}
The main Higgs search channels relevant for the current running
conditions of the LHC will be presented in the following.

\subsection{Higgs to WW$^{*}$}
The H$\rightarrow$WW$^{*}\rightarrow$ll$\nu\nu$ decay channel is considered the 
discovery channel for
a SM Higgs boson in a wide mass range at the LHC. The Higgs to WW$^{*}\cdot$
branching ratio is close to 1 in the 2$\MW < \MH<2\MZ$ mass range and the
leptonic decay of the W bosons gives a clear experimental signature
characterized by the two high p$_{T}$ leptons with opposite charge and a
small transverse opening angle.

Missing transverse energy is also expected, due to the undetected
neutrinos. No central jet activity is characteristic of the 
gg$\rightarrow$H process while two high rapidity jets are expected for
the VBF process (smaller cross-section).

The backgrounds to consider in the analysis are all sources of real
or fakes multi-lepton final states and missing E$_{T}$, like the
irreducible continuum WW production (plus other di-boson processes
such as WZ and ZZ), t$\mathrm{\bar{t}}$ process, Drell-Yan and W+jets, amongst others.

The CMS analysis consider the three final states ee, $\mu\mu$ and e$\mu$.  
As no mass peak can be reconstructed due to the presence of the neutrinos, 
the best knowledge of the backgrounds is mandatory. This
is achieved by using control regions and data-driven
methods. The systematic uncertainties are carefully addressed.

Different approaches have been studied.
A sequential cut-based analysis independently optimized for the three
considered final states in the 0-jet bin (H+0j), a multivariate analysis considering 
all the three final states together and a dedicated study for the VBF
channel.

For an integrated luminosity of 1~$\ifb$ at 7~$\TeVcc$, as shown in Fig.~\ref{WW exclusion} and Fig.~\ref{WW discovery}, exclusion would be possible 
at 95\% CL in this channel from around 140 to 180~$\GeVcc$, 
and discovery level sensitivity (5$\sigma$)
would be achieved for masses around 165~$\GeVcc$.

\begin{figure*}[t]
\centering
\subfloat[]{
\includegraphics[width=100mm]{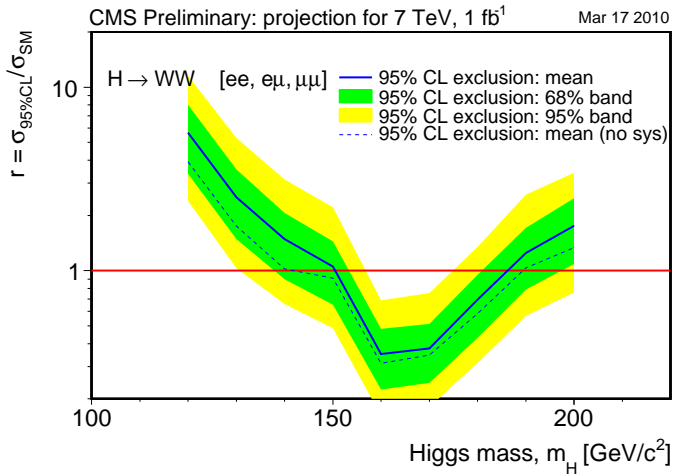}
\label{WW exclusion}
}
\subfloat[]{
\includegraphics[width=70mm]{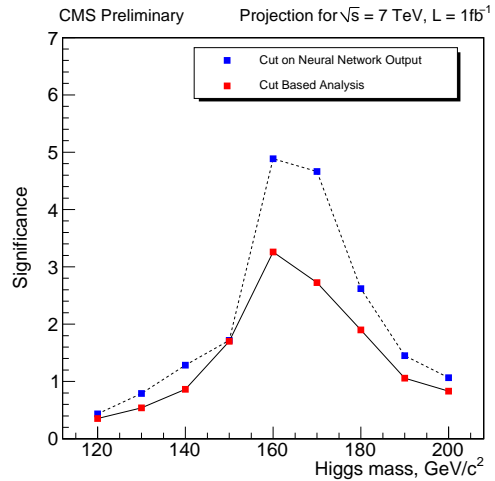}
\label{WW discovery}
}
\caption{Exclusion limit (a) and discovery sensitivity (b) for the Standard Model Higgs boson decay mode WW.
The results are obtained with projections for an integrated luminosity of 1~$\ifb$ at 7~$\TeVcc$.} 
\label{WW puppa}
\end{figure*}

\subsection{Higgs to ZZ$^{*}$}
The H$\rightarrow$ZZ$^{*}\rightarrow$llll decay channel is known as the golden Higgs decay, 
as it has the cleanest experimental signature for discovery with a narrow
four-lepton invariant mass peak on top of a smooth background. It is
powerful in a wide high mass range, however it can be a challenge for
Higgs masses between 120 and 150~$\GeVcc$, where one of the Z bosons is
off-shell.

In the analysis two pairs of same flavor, opposite-sign leptons (4e,
4$\mu$, 2e2$\mu$), are selected and the Higgs mass is reconstructed. 
The main backgrounds are the irreducible ZZ$^{*}$, Zb$\mathrm{\bar{b}}$ and 
t$\mathrm{\bar{t}}$.
Due to the presence of two non-isolated and displaced leptons the last two
can be largely reduced with isolation and impact parameter cuts. 
The rate of the ZZ is assessed from data Z events.

With an integrated luminosity of 1~$\ifb$ at 7~$\TeVcc$, as shown in Fig.~\ref{ZZ exclusion}, 
the SM Higgs boson cannot be excluded anywhere in the entire mass range.

\begin{figure*}[t]
\centering
\includegraphics[width=100mm]{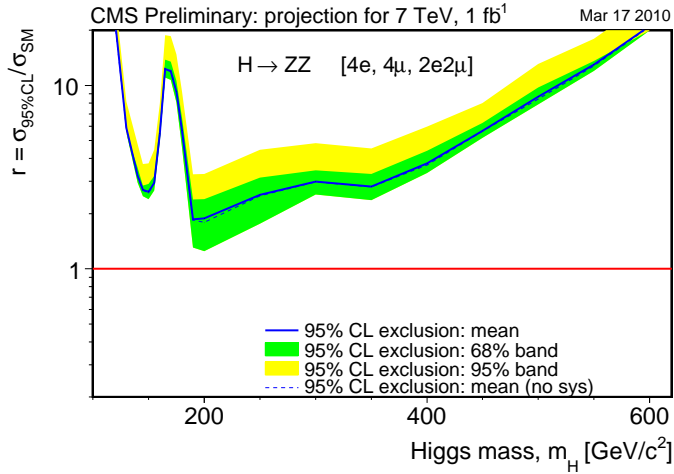}
\caption{Exclusion limit for the Standard Model Higgs boson decay mode ZZ$^{*}$.
The results are obtained with projections for an integrated luminosity of 1~$\ifb$ at 7~$\TeVcc$.} 
\label{ZZ exclusion}
\end{figure*}

However, the Higgs boson with a mass $\MH < 400 \GeVcc$ would be
excluded, if a fourth generation of quarks exists.
Indeed an extra doublet of quarks would make the gluon fusion production rate about  9 times 
larger, regardless of how massive the two extra 4$^{th}$ generation quarks might
be.

\subsection{Higgs to $\gamma\gamma$}
In the low mass range, $110 < \MH < 140 \GeVcc$, H$\rightarrow\gamma\gamma$ is a promising
channel. Two high energy isolated photons in the final state allow
for a mass peak reconstruction, but due to the small branching ratio, this channel
is considered a high luminosity analysis.

The backgrounds are the irreducible $\gamma\gamma$, $\gamma$+jets, jets and
Drell-Yan. The background can be assessed from the sidebands.
CMS has proposed two complementary analyzes: a cut-based and an event-by-event kinematic
likelihood ratio.

For an integrated luminosity of 1~$\ifb$ at 7~$\TeVcc$ as shown in Fig.~\ref{gammagamma exclusion} the SM Higgs boson cannot be excluded anywhere in the entire mass range.

\begin{figure*}[t]
\centering
\includegraphics[width=100mm]{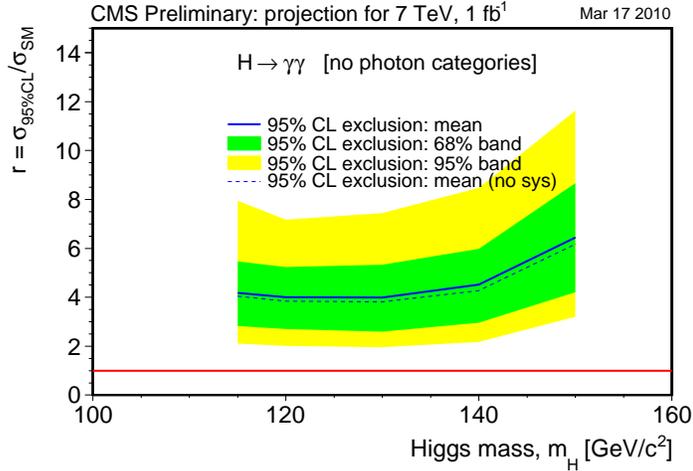}
\caption{Exclusion limit for the Standard Model Higgs boson decay mode $\gamma\gamma$.
The results are obtained with projections for an integrated luminosity of 1~$\ifb$ at 7~$\TeVcc$.} 
\label{gammagamma exclusion}
\end{figure*}

The case of a fermiophobic Higgs would largely suppress the gluon fusion cross
section enhancing at the same time the $\gamma\gamma$ decay mode.
Under this condition and assuming a conservative selection it would be possible 
to improve the SM exclusion limit by a factor four or down to 110~$\GeVcc$.

Fig.~\ref{combined exclusion} shows the projected exclusion limits for a SM Higgs by
combining results of the H$\rightarrow$WW$^{*}$, H$\rightarrow$ZZ$^{*}$, and H$\rightarrow\gamma\gamma$ channels. 
The expected exclusion mass range is $145 < \MH <190 \GeVcc$. 
The Higgs boson with a mass $\MH < 500 \GeVcc$ would be excluded, if a
fourth generation of heavy quarks exists.

\begin{figure*}[t]
\centering
\includegraphics[width=100mm]{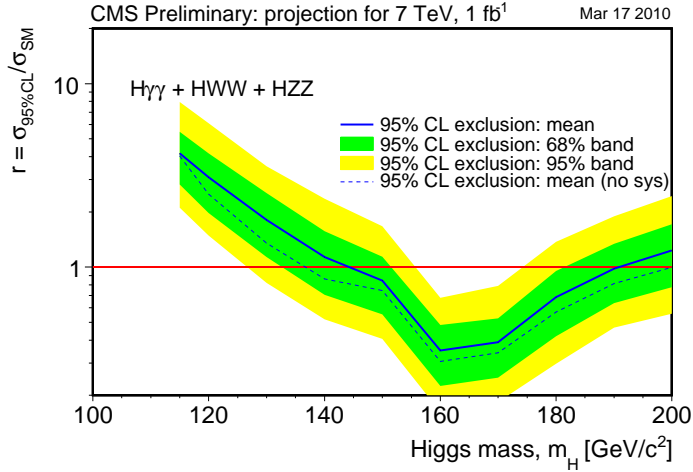}
\caption{Combined exclusion limit for the Standard Model Higgs boson.
The results are obtained combining the results of the projections for an integrated luminosity of 1~$\ifb$ at 7~$\TeVcc$ of three decay modes: WW, ZZ$^{*}$ and $\gamma\gamma$.} 
\label{combined exclusion}
\end{figure*}

\section{MSSM HIGSS SEARCHES CHANNEL AT THE LHC}
The main channel for minimal Super Symmetric Standard Model (MSSM)
Higgs searches in the first period of running of the LHC will be heavy
neutral MSSM Higgs $\Phi$ produced via b quark association with subsequent
decay to $\tau$, b$\mathrm{\bar{b}}\Phi$, $\Phi\rightarrow\tau\tau$.

The main backgrounds are Z+b$\mathrm{\bar{b}}$/c$\mathrm{\bar{c}}$/jets and 
t$\mathrm{\bar{t}}$. 
In the proposed analysis final states with isolated pairs of $\tau$ 
decaying to hadrons and leptons are selected.
Missing E$_{T}$ is also required as well as one b-tagged jet and a veto on 
extra jets in the event.

The analysis is performed by counting events in the $\tau\tau$ invariant
mass window, reconstructed by collinear approximation where the $\tau$
decay products are assumed to be in the same direction as the $\tau$ itself.

Fig.~\ref{mssm discovery} shows the projected discovery and exclusion contours in the
MSSM  ($\MA$, tan$\beta$-plane) for the search for a neutral MSSM Higgs
bosons in the pp$\rightarrow$b$\mathrm{\bar{b}}\Phi$, b$\mathrm{\bar{b}}\tau\tau$ channel 
with the CMS experiment. 
At $\MA ~ 90 \GeVcc$, discovery is possible for tan$\beta > 20$ and the
exclusion limit is expected to reach down to tan$\beta >15$.

\begin{figure*}[t]
\centering
\includegraphics[width=100mm]{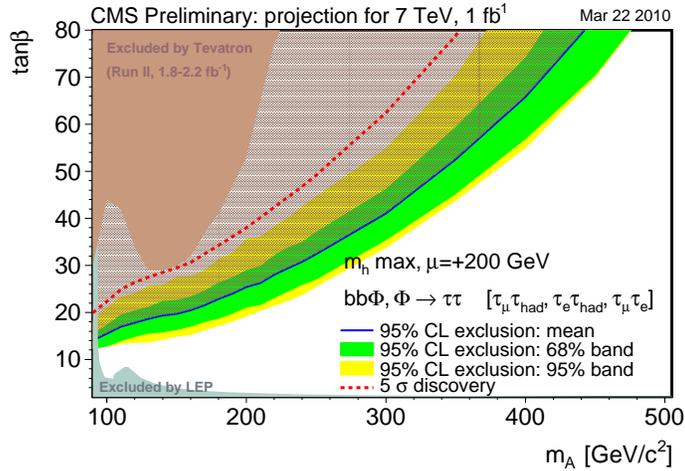}
\caption{Exclusion and discovery contours for a neutral MSSM Higgs boson in the 
pp$\rightarrow$b$\mathrm{\bar{b}}\Phi$, b$\mathrm{\bar{b}}\tau\tau$ channel.
The results are obtained with projections for an integrated luminosity of 1~$\ifb$ at 7~$\TeVcc$.} 
\label{mssm discovery}
\end{figure*}

\section{CONCLUSIONS}
The performance of the CMS detector in collision data has been very
good since the beginning of the data taking allowing to produce the 
first results after few hours. 

The reconstruction of the main physics objects used in the Higgs analysis 
is performing well and it is already possible to start exploring Standard Model
processes, namely W and Z.

At 7~$\TeVcc$, with enough luminosity (1~$\ifb$), the CMS experiment will
begin to explore a sizable range of Higgs mass, reaching SM Higgs
discovery sensitivity for masses between 160 and 170~$\GeVcc$ and
exclusion between 140 and 200~$\GeVcc$, while low mass SM Higgs searched
will require higher center of mass energy and integrated luminosity.

For MSSM Neutral Higgs the discovery range at $\sqrt{s} = 7 \TeVcc$ for
small $\MA$ would be tan$\beta > 20$ and exclusion be possible down to
tan$\beta~15$.

\end{document}